\documentclass[10pt,journal]{IEEEtranTCOM}

\usepackage[english]{babel}
\usepackage[usenames]{color}
\usepackage[cp1250]{inputenc}
\usepackage{amsfonts}
\usepackage{amssymb}
\usepackage{amsthm}
\usepackage{graphicx}
\usepackage{epsfig}
\usepackage{mathrsfs}
\usepackage{amsmath}
\usepackage{algorithm}
\usepackage{algorithmic}
\usepackage{hyperref}
\usepackage{mathtools}

\theoremstyle{plain}

\begin{document}
\newcommand{\bea}{\begin{eqnarray}}
\newcommand{\eea}{\end{eqnarray}}
\newcommand{\be}{\begin{equation}}
\newcommand{\ee}{\end{equation}}
\newcommand{\beas}{\begin{eqnarray*}}
\newcommand{\eeas}{\end{eqnarray*}}
\newcommand{\bs}{\backslash}
\newcommand{\bc}{\begin{center}}
\newcommand{\ec}{\end{center}}
\def\SC {\mathscr{C}}

\title{{Testing stimulated emission photon directions}}
\author{\IEEEauthorblockN{Jarek Duda}\\
\IEEEauthorblockA{Jagiellonian University, Krakow, Poland, \emph{jaroslaw.duda@uj.edu.pl}}}
\maketitle

\begin{abstract}
While naively laser only causes excitation of external target, e.g. Rabi cycle, STED microscopy or ASE/SASE/SSA demonstrate it can also stimulate its deexitation, however, under uncommon condition of being prepared as excited. These two causalities are governed by absorption-stimulated emission pair of equations, and swap places in perspective of T/CPT symmetry, however, it means photon direction of stimulated emission should be opposite to usually assumed, allowing for negative radiation pressure $\vec{p}=\langle \vec{E}\times \vec{H}\rangle/c$. This article discusses various arguments and proposes simple direct tests to experimentally verify existence of such backward photon trajectories, complementing consequent forward textbook trajectories. Depending on the results, it could lead to many proposed applications like medical, astronomical or 2WQC more symmetric quantum computers. Alternatively, if unsuccessful, it would require macroscopic violation of CPT symmetry, so far tested probably only in microscopic settings.
\end{abstract}
\textbf{Keywords:} lasers, photonics, STED microscopy, stimulated emission, ASE/SASE/SSA, reverse bias, negative radiation pressure, CPT symmetry, ring laser, synchrotron radiation, 2WQC, optical fiber, EDFA amplifier, optical isolator
\section{Introduction}

\begin{figure}[b!]
    \centering
        \includegraphics[width=9cm]{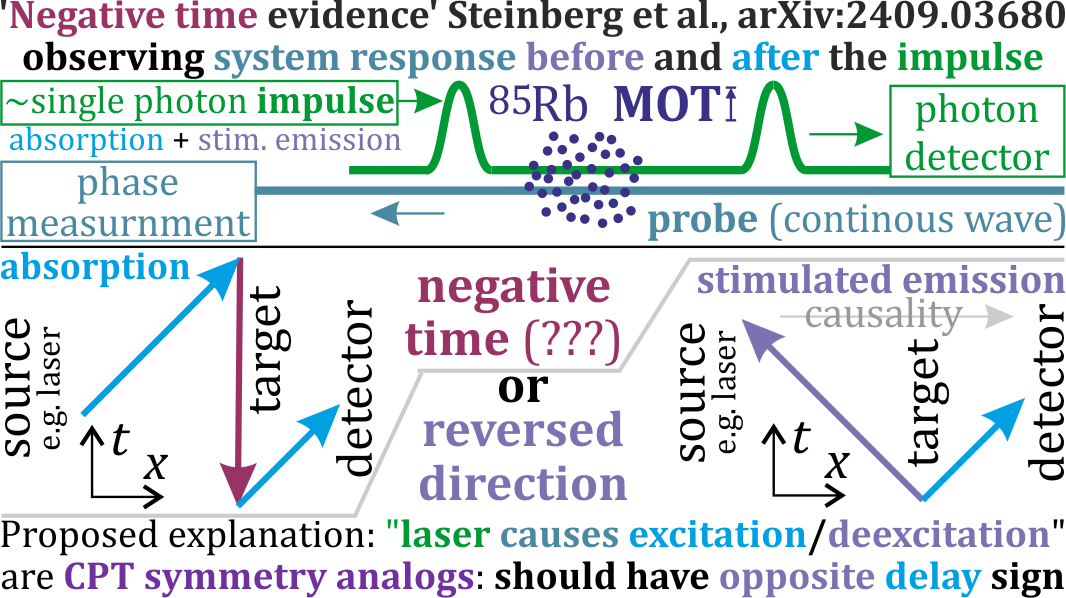}
        \caption{\textbf{Top}: recent widely commented experiment by Steinberg group~(\cite{backward, negative}) has observed medium response (measured phase of \textsuperscript{85}Rb) before and after the applied impulse, proposing \textbf{"photon can spend a negative amount of time in an atom cloud"} explanation, which does not seem physical, however, theoretical explanation for this article uses forward/backward-in-time propagators~\cite{negexp}.\\ \textbf{Bottom}: as discussed here, intuitive explanation of such forward/backward propagators, like in two-state vector formalism~\cite{tsvf}, could come from \textbf{absorption/stimulated emission} being CPT analogs, hence should have opposite delay. 
        }
        \label{negative}
\end{figure}

\begin{figure}[t!]
    \centering
        \includegraphics[width=9cm]{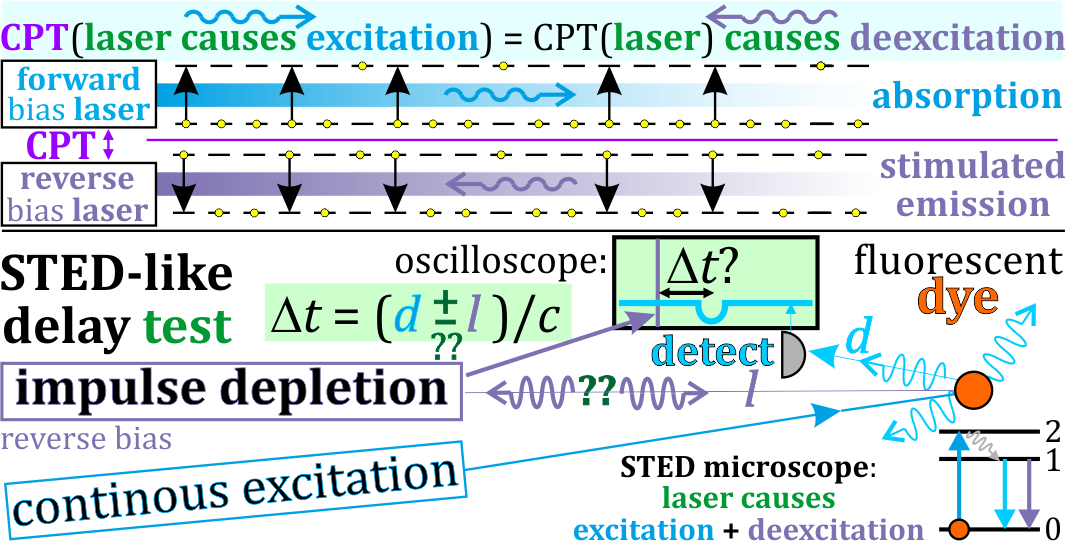}
        \caption{\textbf{Top}: CPT symmetry says that if there exists "laser causes excitation" scenario, also "laser causes deexciation" should exist (needs reversing bias: voltage). Being CPT symmetry analogs, they should have opposite photon direction and delay sign - we would like to verify it experimentally and apply if successful. For example in \href{https://en.wikipedia.org/wiki/STED_microscopy}{STED microscopy} diode lasers cause both. \\
        \textbf{Bottom}: proposed STED-like test of this photon direction(s): continuous laser excites fluorescent dye being approximately 3-level medium, and impulse laser causes its deexcitation through stimulated emission - its photon direction can be found from $\Delta t=(d\pm l)/c$ delay between laser impulse and observed reduced intensity by fast photodetector focused on dye spontaneous emission. Textbooks assumption would mean $'+'$ sign here, while CPT symmetry requires $'-'$ sign instead, hence this is also macroscopic test of CPT symmetry, complementing dozes of earlier microscopic tests~\cite{CPTdata}. Applying reverse bias for depletion laser as in Fig. \ref{negtemp} brings hope for STED avoiding photobleaching. }
        \label{testmin}
\end{figure}

\begin{figure}[t!]
    \centering
        \includegraphics[width=9cm]{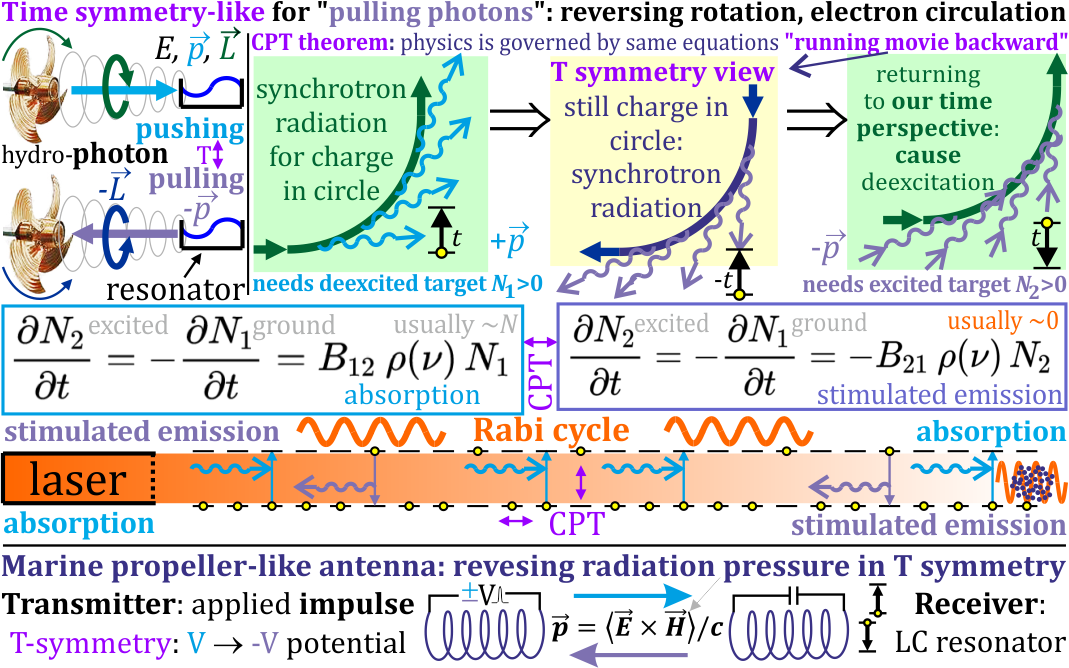}
        \caption{\textbf{Top}: swirl-like wave behind marine propeller carries energy, momentum, and angular momentum like photon, also allowing to perform time-symmetry-like transformation by just reversing rotation, getting "pulling" photons able to pull energy from excited resonator. EM is mathematically close to hydrodynamics (e.g. \cite{EMh}), suggesting to analogously search for EM "pulling photons" using e.g. synchrotron radiation: coming from just circulating electron - we could also reverse by just changing electron direction, what should analogously reverse $\vec{p}=\langle \vec{E}\times \vec{H}\rangle/c$
        radiation pressure. T/CPT symmetry also switches shown absorption and stimulated emission equations, allowing to estimate strength of the effect. 
        Practically it could be realized e.g. by betatron, synchrotron, free electron laser (FEL) with target positioned for reversed electron trajectory. \textbf{Bottom}: for lower frequencies there might be used analogous to marine propeller e.g. spring-like antenna powered with impulse: reversing $\vec{p}=\langle \vec{E}\times \vec{H}\rangle/c$ radiation pressure for reversed time ($t\to -t, H\to -H$), or used impulse electric potential $V\to -V$ - hopefully allowing both to excite resonator of receiver as in wireless charging, but also speedup it deexcitation/relaxation for reversed impulse.  }
        \label{synch}
\end{figure}

CPT symmetry (of charge conjugation (C) + parity transformation (P) + time reversal (T)) is shown~\cite{CPT} to be necessary for local Lorentz invariant quantum field theories, hence is generally believed to be satisfied by equations governing our nature, working analogously if running time backward. Moreover, electromagnetism we focus on here, is already T-symmetric, explored e.g. in Wheeler-Feynman absorber theory~\cite{WF} using retarded and advancing waves. In contrast, 2nd law of thermodynamics is asymmetric, however, it is only statistics of solution we live in, like throwing a rock to lake surface, which is symmetric in equations. Arguments for entropy growth like Boltzmann's H theorem~\cite{bH} use mean-field-like approximation called Stosszahlansatz. Applying CPT symmetry before such proof, we would get reversed entropy growth, well summarized in \cite{emergence} as "the system is dissipative and decohering in both temporal directions".

CPT symmetry has dozens of successful experimental confirmations~\cite{CPTdata}, however, it seems all tests so far were microscopic - come from observations of single particles. Conscious macroscopic test was not found in literature, important especially due to T time symmetry which might lead to new applications - some are proposed and discussed here. For example by acting only with stimulated emission on target, negative radiation pressure. 

\begin{figure}[t!]
    \centering
        \includegraphics[width=9cm]{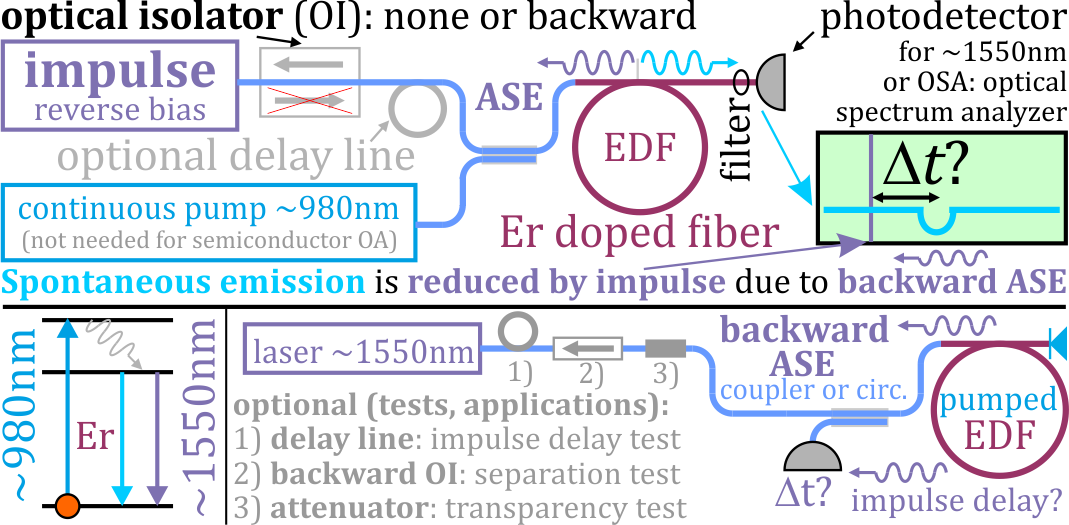}
        \caption{In commercially available e.g. erbium-doped optical fiber amplifiers (EDFA, or simpler semiconductor SOA, etc.)~\cite{erb} there is used backward ASE (amplified spontaneous emission): increased emission from pumped doped fiber, also toward laser e.g. in ~\cite{bASE}, exactly as in the sought CPT analog of "laser causes excitation" effect, usually removed by forward optical isolators. \textbf{Top}: Analogously to STED-like test, we can verify this photon direction by measuring delay between the impulse and reduced population of excited atoms of external target, e.g. by monitoring its spontaneous emission. For this purpose, the shown standard forward optical isolator should be removed, or rotated to backward position. \textbf{Bottom}: simplified setting from \cite{bASE} seems much more practical for  tests: of delay (opposite if CPT analog), separation (of forward/backward action), transparency (forward blocked by $N_1$, so backward should be by usually much lower $N_2$) - allowing for potential future applications. As in Fig. \ref{negtemp}, it seems crucial to use reverse bias for the laser, in practice (e.g. \cite{backward,negative,bASE,superr}) probably included in varying/impulse signal. }
        \label{erb}
\end{figure}

Applying CPT symmetry to "laser causes target excitation" scenario, it becomes "CPT(laser) causes CPT(target) deexcitation", replacing absorption equation with stimulated emission, having photons traveling from CPT(target) to CPT(laser), also reversing time delay, what might explain observations of negative time delays e.g. by Steinberg group as in Fig. \ref{negative}~\cite{backward,negative}, also system response before impulse like in \cite{superluminal}, finally being a consequence of large number of single emissions/absorptions. 

We would like to verify existence of such CPT analog of "laser causes excitation" scenario with backward photon trajectory, for example in STED-like setting as in Fig. \ref{testmin}. Well known backward ASE (amplified spontaneous emission) in common optical amplifiers, as in Fig. \ref{erb}, seems exactly the sought scenario, also providing practical realization of such test, and maybe also discussed further potential applications thanks to wide commercial availability of optical fiber hardware. Its synchrotron radiation analogs are SASE (self-amplified spontaneous emission)~\cite{SASE} used in FEL~\cite{FEL}, and synchrotron self-absorption (SSA) considered in astrophysics~\cite{SSA}.

\begin{figure}[t!]
    \centering
        \includegraphics[width=9cm]{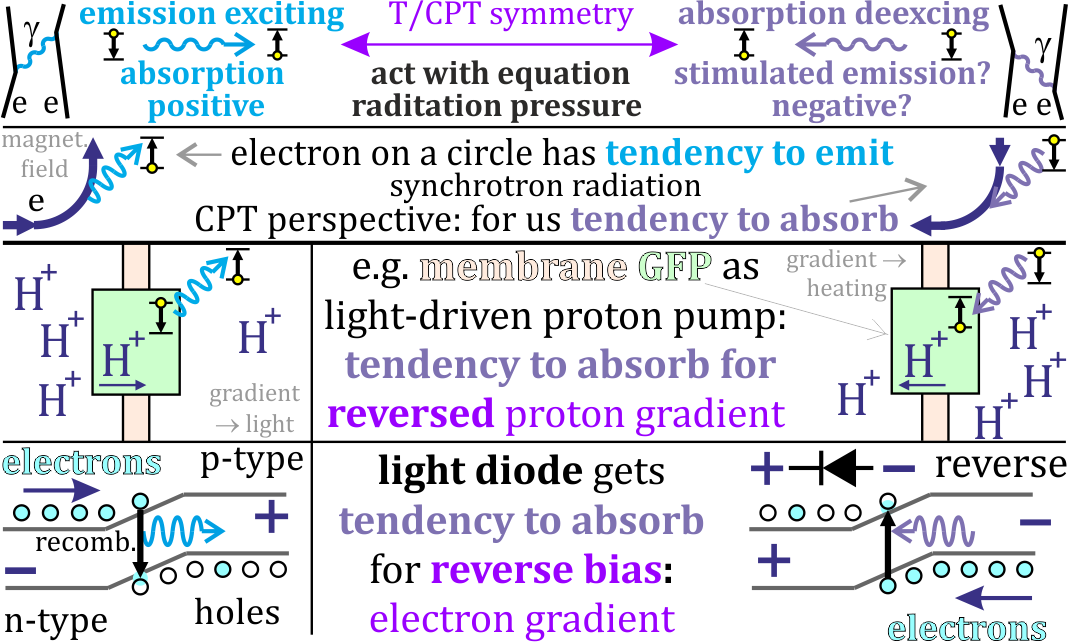}
        \caption{Usually system has tendency to get to the ground state, emitting photon, which e.g. excites some atom, according to absorption equation. However, getting reversed tendency should increase amplitude (probability) of shown symmetric reversed Feynman diagram: causing absorption of photon, emitted by some atom - according to symmetric stimulated emission equation.  Beside previous based on synchrotron radiation, diagram shows two examples of obtaining such revered tendency: upper uses membrane GFP (green fluorescent protein) which can act as light-driven protein pump~\cite{gfp}, in reversed proton gradient would get tendency to absorb photons, what might be used e.g. by biological organisms for thermoregulation. The bottom is analogously applied reversed bias/potential/electron gradient for light/laser diode, what is also applied for light detectors\cite{rLED}, but should actually also stimulate emission - e.g. used as absorber in \cite{superr} clearly helps with generation of photon impulse. Impulse laser sources often have e.g. sinc-like shape, so should also include reversed bias, what might explain observations in \cite{backward, negative}. Backward ASE in \cite{bASE} probably also uses both forward and reverse bias in varying signal, if successful would be perfect for presented applications. Also e.g. Fig. \ref{testmin}, \ref{erb} tests should rather use reverse bias. Alternatively, change of $N_2$ above/below thermal $N_2 \sim N_1 \exp(-\beta (E_2-E_1))$ of Boltzmann distribution should also increase amplitude of Feynman diagram with emission/absorption there, and absortpion/emission in the target.
        }
        \label{negtemp}
\end{figure}

In contrast, as in Fig. \ref{test}, standard textbook view on stimulated emission is energy being kind of knocked forward by incoming photon - with the opposite direction for the emitted photon. There are lots of experimental evidence for existence of such process. While it seems in disagreement with CPT symmetry, bottom of Fig. \ref{test} shows it can be viewed as a consequence of the sought: CPT analog of "laser causes excitation" scenario. Radiation pressure seems a good intuition - usually positive stimulating emission outward, but there should be also symmetric: stimulating emission inward, like marine propeller pushing/pulling energy from resonator.

Simple tests with continuous diode laser were performed, but without success. This version of article suggests it was because of using constant forward bias, while as in Fig. \ref{negtemp} it seems crucial to use reverse bias to reverse emission tendency to absorption. Successful realization applied varying signal, impulse lasers instead~(\cite{backward, negative,superluminal}), which seem to contain reverse bias contributions. It is planned to be tested soon with reverse bias.

In contrast, promising simpler situations are offered by synchrotron radiation and antennas, as shown in Fig. \ref{synch}. Synchrotron radiation is emitted by e.g. electron travelling on a circle, being also so from CPT perspective. Related tests were earlier proposed using free electron laser~\cite{my} (\href{https://web.archive.org/web/20230617055541/https://groups.google.com/g/sci.physics.foundations/c/xhUfe8akaS0/m/9M80Fvsc-q4J}{earlier 2009 post}). However, organization of such test has turn out also difficult.

Tests on antennas seem in reach, hopefully to be performed in near future. They are inspired by analogy between electromagnetism and hydrodynamics (e.g. \cite{EMh}, or Barnett effect~\cite{barnett}) - marine propeller is asymmetric, creating both positive and negative pressure, switched by reversing rotation/applying T symmetry - it could remotely excite/deexcite resonator. E.g. spring-like antenna should have analogous asymmetry, switching $\vec{p}=\langle \vec{E}\times \vec{H}\rangle/c$ radiations pressure for T-symmetry reversing magnetic field $H$. We should try to remotely excite resonator this way by applying impulse to such transmitter antenna, and try to speedup its deexcitation/relaxation by applying opposite impulse.

\begin{figure}[t!]
    \centering
        \includegraphics[width=9cm]{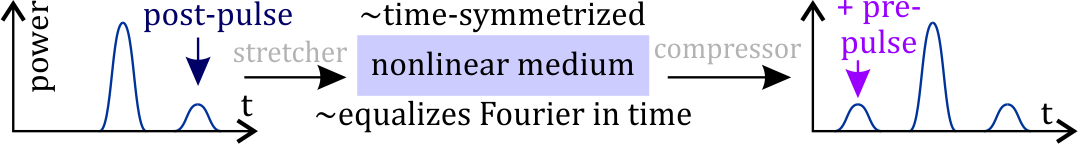}
        \caption{Other type of argument for time/CPT symmetry of lasers are pre-pulses generated by post-pulses, observed (e.g. \cite{prepulse0,prepulse}) for extremely high energy lasers based on chirped-pulse amplification (CPA) going through nonlinear medium (being first stretched then compressed). Theoretical analysis is based on Fourier transform in time, being symmetrized by nonlinearities. However, decomposing it into individual absorptions/emissions, negative delays should again come from being CPT analogs, hence having opposite delay signs.}
        \label{prepulse}
\end{figure}
\begin{figure}[t!]
    \centering
        \includegraphics[width=9cm]{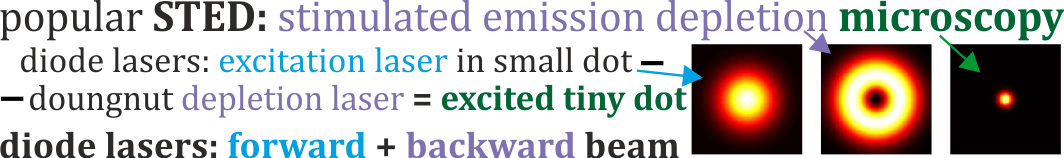}
        \caption{
        \href{https://en.wikipedia.org/wiki/STED_microscopy}{STED microscopy (image source)} is a popular application of stimulated emission - using additional laser reducing excitation in a doughnut shape, leaving smaller dot to increase resolution.}
        \label{STED}
\end{figure}

\section{STED microscopy based tests}
Stimulated emission depletion (STED) microscopy~\cite{STED} improves resolution by reducing size of excited dot with donought-like depletion beam from second laser as in Fig. \ref{dirs}. It uses fluorescent dye being approximately 3 level medium: excited with one laser from 0 to 2, then fast relaxation to 1, then deexcitation from 1 to 0 with slightly longer wavelength - usually through just spontaneous emission observed by microscope lens. 

In STED we additionally speedup this deexcitation through stimulated emission by second: depletion laser of frequency for $1\to 0$ transition - actively reducing the number of photons from doughnut-shaped region. This way the final excited dot is smaller, increasing microscope resolution for lens focused on stimulated emission of the dye.

\begin{figure}[t!]
    \centering
        \includegraphics[width=9cm]{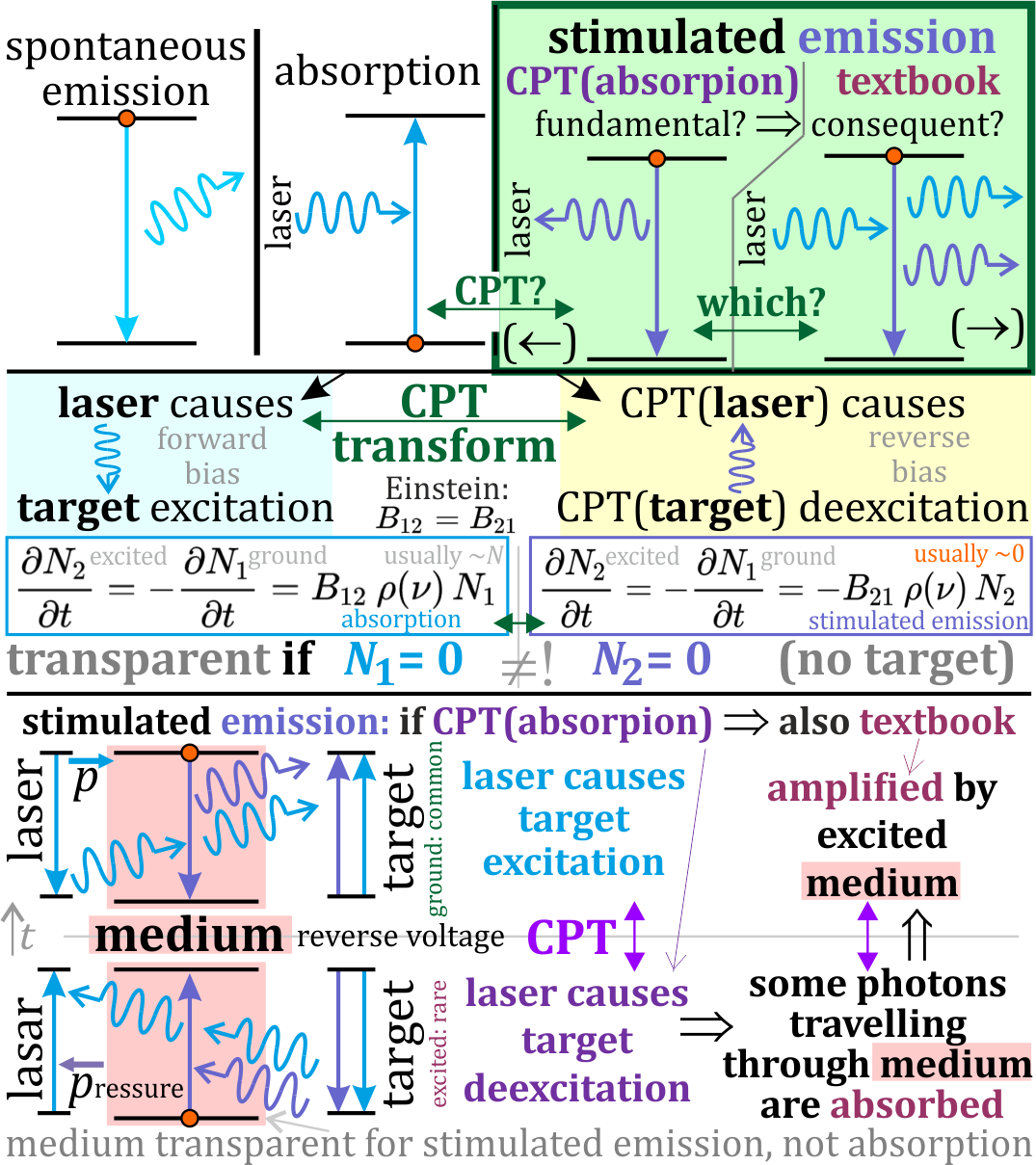}
        \caption{\textbf{Top}: "excited atom $\rightleftarrows$ deexcited + photon" scenarios: spontaneous emission in random direction, direction dependent: absorption and \textbf{stimulated emission} in focus: in textbook $(\rightarrow)$ forward view incoming photon kind of knocks out energy (in backward ASE in opposite direction), while CPT symmetry suggests backward: laser $(\leftarrow)$ target instead. The latter allows to conclude the former, hence the question to verify is existence of CPT(absorption) effect.\\ 
        \textbf{Center}: basic argument for $(\leftarrow)$ is CPT symmetry of physics: transforming "laser causes target excitation" scenario into (lasar=) "CPT(laser) causes CPT(target) deexcitation", requiring to reverse voltage (bias) as in Fig. \ref{negtemp}, switching the stimulated emission-absorption equations, governed by the same (CPT symmetric) Einstein's coefficients $B_{12}=B_{21}$. \\
        \textbf{Bottom}: schematic explanation that CPT view $(\leftarrow)$ also allows to conclude textbook $(\rightarrow)$ picture, hence both can be true. For photons travelling in 
        CPT analogous scenarios (using reversed bias: voltage), we add a medium on the way: looking at transparency in the latter, stimulated emission should fully act on the target, but some photons it produces should be absorbed by medium - returning from CPT perspective it becomes textbook amplification by going through excited medium. Notice asymmetry for target (also in further Fig. \ref{asym}): it is now easy to find ground state target, but difficult for excited - making the upper scenario more likely. It is natural to think about it as acting with positive/negative radiation pressure: stimulating outward/inward photon emission (if possible), blocked by deexcited/excited targets on the way. }
        \label{test}
\end{figure}

While spontaneous emission photons should have a random direction, \textbf{the question of concern is direction of stimulated emission photons: $(\rightarrow)$ forward ano/or $(\leftarrow)$ backward?} Textbooks assume forward, discussion here based on CPT symmetry suggests backward as fundamental, and forward as its consequence (Fig. \ref{test}). Therefore, both can be true, the question to verify is existence of backward $(\leftarrow)$. All tests require laser without built-in forward optical isolator, and rather powered by reversed bias.

\subsection{Delay STED-like test (needs reverse bias)}
Assuming there are only these two options: $(\leftarrow)$ or $(\rightarrow)$ and we would like to test existence of the former, the simplest way to distinguish them seems delay. There is literature discussing delay in STED microscopy, like \cite{STED1,STED2,STED3}, however, data they provide seems insufficient to uniquely answer this question.

Therefore, there is proposed STED-like setting as in the bottom of Fig. \ref{testmin}: two (e.g. diode) lasers: continuous for $0\to 2$ excitation of the dye, and impulse depletion for its  $1\to 0$ stimulated emission. There is also photodetector observing spontaneous emission from this dye, which should observe reduced intensity due to depletion laser impulse - the question is its delay. 

Hence (fast) photodetector should be connected to oscilloscope, triggered by depletion laser impulse. Its observed delay $\Delta t$ contains positive contribution $d/c$ for light path from dye to detector, and $\pm l/c$ contribution from impulse laser-dye path, of sign for one of the two assumed possibilities - which could be determined from this delay e.g. with oscilloscope snapshots.

Fluorescent dye can be replaced with a different medium for which we can maintain a relatively high population level of excited atoms, what is much more accessible for 3,4-level media. Impulse preferably should have a fraction of nanosecond, (e.g. cascade) photodetector needs to allow nanosecond-scale time resolution. However, their imperfections could be compensated by increasing $l$ distance to impulse laser.

As in Fig. \ref{erb}, standard commercially available EDFA \textbf{fiber optics} equipment seems very convenient for analogous test and further applications if successful (suggested by Micha\l{} Markiewicz). The sought target $\to$ laser photon trajectories caused by laser seems exactly the backward ASE effect, well known in literature - usually unwanted and removed by optical isolator. We can test these photon trajectories by removing or better reversing this optical isolator, adding (rather fast) photodetector filtered to observed the stimulated emission, and observing delay of its reduction as response for impulse of deexcitation laser. In comparison to STED setting, $d/c$ propagation time is between EDF and detector, $l/c$ between EDF and impulse laser - easily extended by delay line.
\subsection{Alternative STED-like tests (needs reverse bias)}
Another approach is placing photodetector in depletion laser-dye line behind the dye, allowing to determine $(\leftarrow)$ or $(\rightarrow)$ from intensities. Measuring also angle dependence, we could also test (rather exclude) change of angle hypothesis.

\begin{figure}[t!]
    \centering
        \includegraphics[width=9cm]{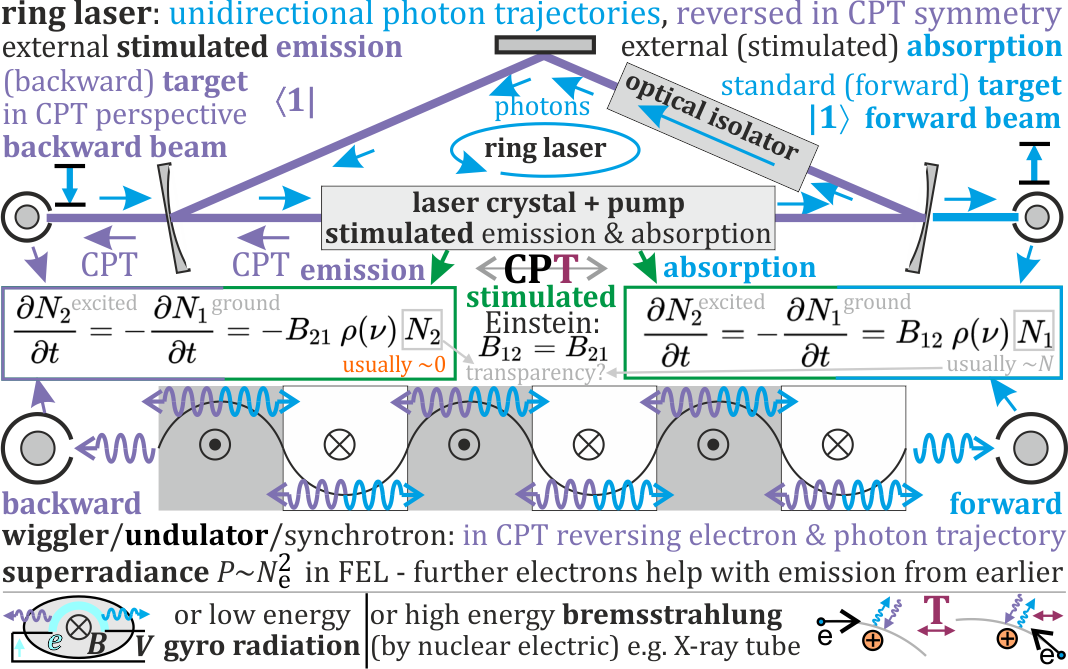}
        \caption{Ring laser uses optical isolator to enforce unidirectional photon circulation, which would be reversed in CPT perspective for reversed bias - switching the action of absorption and stimulated emission equations, suggesting only one of them should act on external targets - let us refer to them as \textbf{forward/backward beam}.      
        Analogously for reverse biased lasers, or synchrotron sources like synchrotron, betatron, wiggler, undulator, free electron laser, bremsstrahlung - reversing electron and photon trajectory/tendency in CPT perspective. FEL reaches superradiance (power $P$ growing with square of the number of electrons $N_e^2$), better coherence, narrower spectrum and  angle~\cite{FEL} - what already seems a consequence of discussed CPT analog of emission: that further electrons stimulate emission from earlier, as in T/CPT view there would be reversed photon exchange: further electrons exciting earlier.}
        \label{dirs}
\end{figure}

\begin{figure}[t!]
    \centering
        \includegraphics[width=9cm]{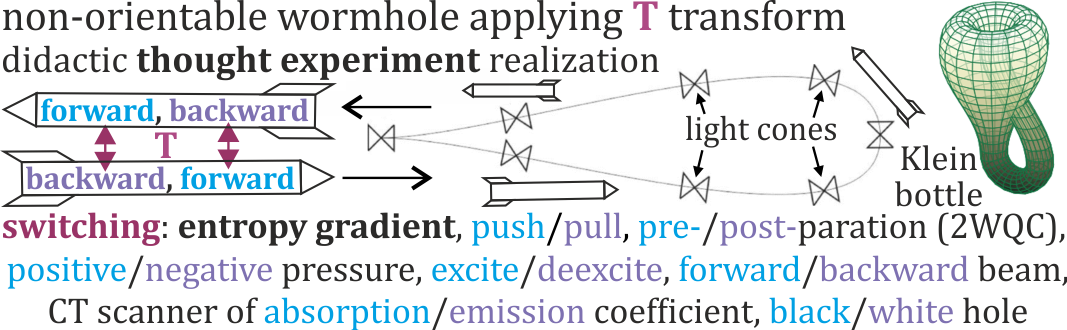}
        \caption{General relativity below black hole horizon rotates time to space direction, in theory could also twice further  e.g. in Klein-botle like \href{https://en.wikipedia.org/wiki/Non-orientable_wormhole}{nonorientable wormhole}~\cite{nonorientable}: applying time symmetry to a rocket travelling through it, getting \textbf{thought experiment} valuable to gain intuitions about time symmetry. For example entropy gradient inside would be reversed. State preparation there would be postparation, allowing for 2WQC~\cite{2WQC} more symmetric quantum computers (Fig. \ref{2wqc}). Forward beam from such rocket would be backward beam for external observer, switching positive and negative radiation pressure. CT scanner there used on external object would map emission coefficient instead of absorption (Fig. \ref{CAT}). Our black holes for observer inside would be white holes (Fig. \ref{BH}).
        }
        \label{nonorient}
\end{figure}

As discussed further, additional tests could use unidirectional e.g. ring laser (or standard with added optical isolator): its forward beam should not allow for depletion here, its backward beam should allow - what could be also tested in the future. 

We could also test suggestions of better transparency for backward beam, e.g. by placing (non-excited) obstacle on the way to depletion laser in STED-like setting. It also brings hope for effects on excited targets already for shielded synchrotron sources. Even if such photons could not cross the obstacle, this effect could be used e.g. for information transfer by trying to detect such additional directional photons from excited target behind obstacle, or monitoring its population level. 

Rabi cycle could be used instead, by applying impulse laser and monitoring/analyzing situation at impulse boundaries, or including optical isolator on the way.

\begin{figure}[t!]
    \centering
        \includegraphics[width=9cm]{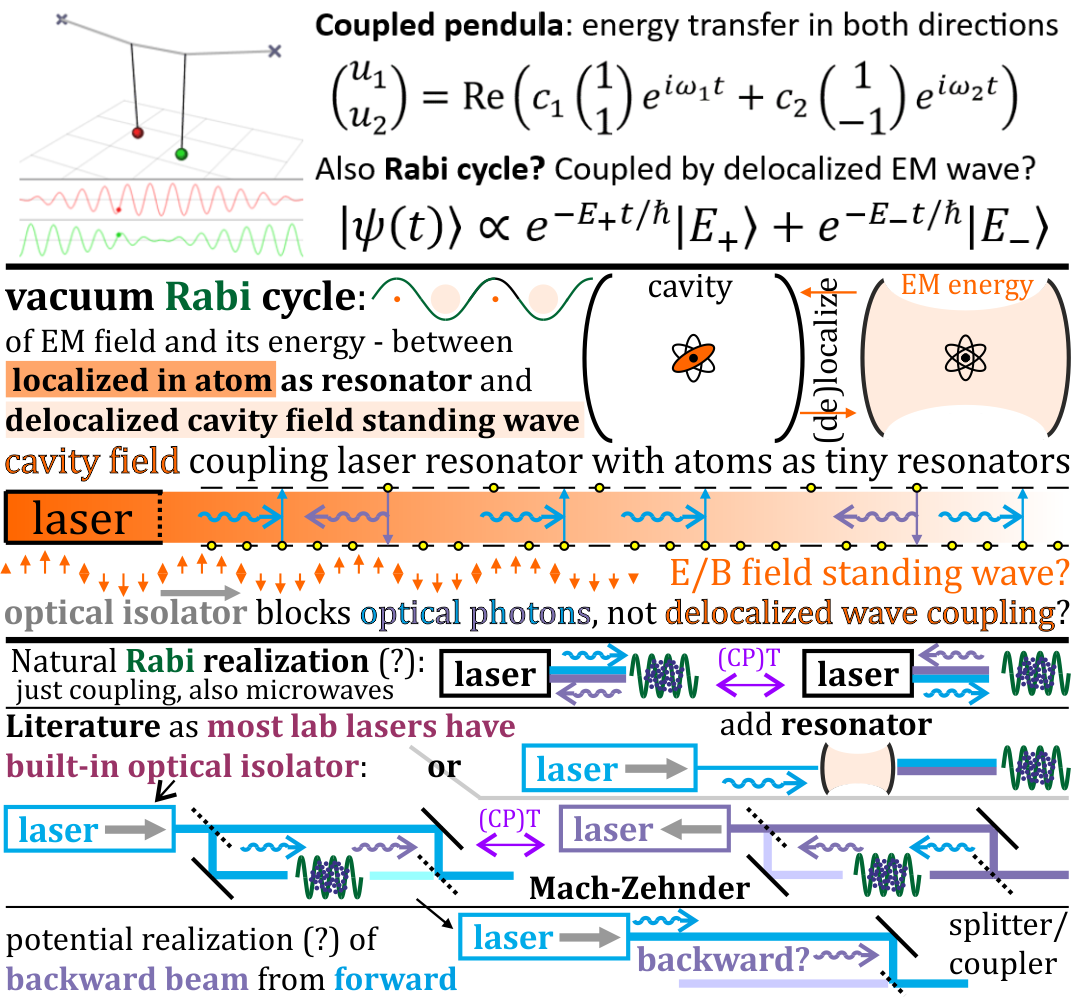}
        \caption{\textbf{Top}:\href{https://en.wikipedia.org/wiki/Pendulum\#Coupled_pendulums}{coupled pendula (image source)} as analog of Rabi cycle (e.g. \cite{crabi1,crabi2,crabi3,crabi4}) - to view it as coupling of e.g. laser as pumped directional resonator, and atom as tiny isotropic resonator, or standing waves of optical cavity as resonator e.g. through  Jaynes-Cummings model~\cite{JC}.\\
        \textbf{Center}: vacuum Rabi cycle~\cite{cavity} - observed oscillations between optical cavity resonator, and atom as tiny resonator - of energy between being localized in atom (high energy density), and delocalized standing EM wave (low energy density) - having different EM fields, like localized laser vs delocalized radio communication.
        Laser can enforce Rabi oscillations, periodically causing target excitation and deexcitation - allowing to view these causalities as results of coupling of two resonators.
        Being able to add some asymmetry to such energy transfer caused by coupling, as discussed further can lead to many applications - e.g. optical isolator brings hope for that: designed to block localized optical photons in one direction, what not necessarily has to be true for delocalized standing wave for coupling - having lower energy density.\\
        \textbf{Bottom}: turns out nearly all lasers have built-in optical isolator, what might prevent required Rabi bidirectional energy exchange like for FEL. Searching literature, while for microwaves Rabi cycle has straightforward realization, for lasers there are used tricks like added resonator~\cite{rl3}, or Mach-Zehnder type setting (\cite{rl1,rl2}), requiring analysis like in "Asking photons where they have been"~\cite{asking}. Just removing this isolator might allow for Rabi simpler realizations. The Mach-Zehnder setting also suggests (to be tested) how to get backward beam having laser with built-in optical isolator: by just adding splitter/coupler.}
        \label{coupling}
\end{figure}

\begin{figure}[t!]
    \centering
        \includegraphics[width=9cm]{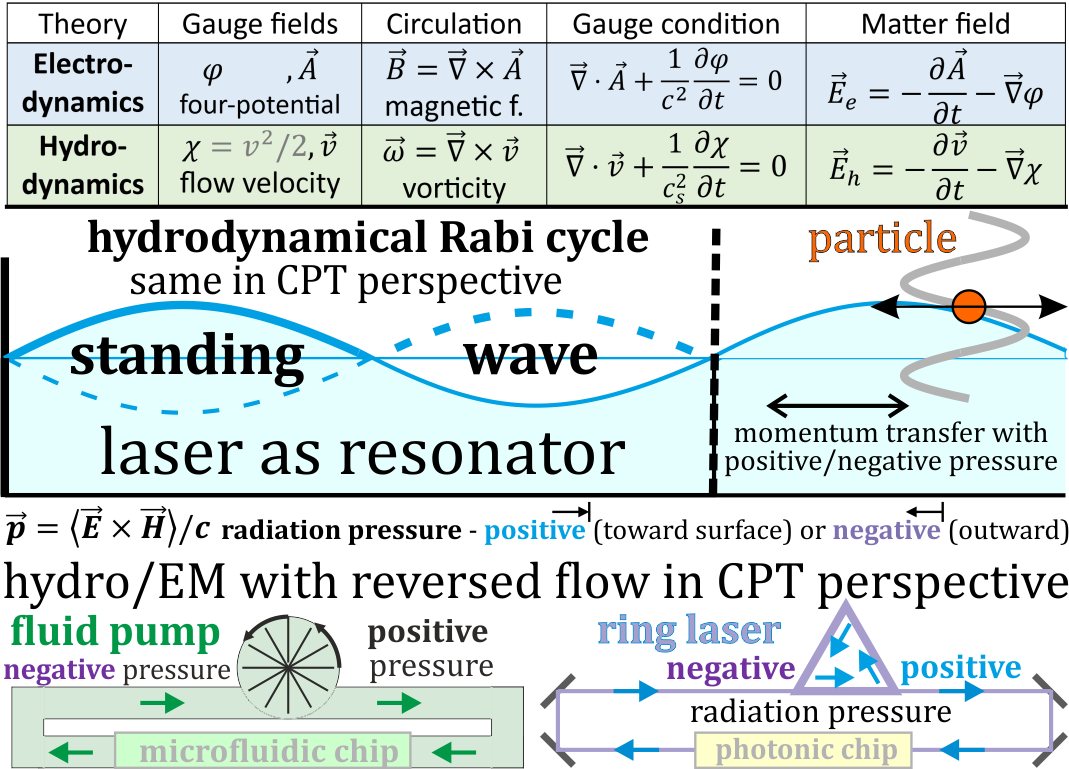}
        \caption{\textbf{Top}: Mathematical analogy between electrodynamics and hydrodynamics (superfluid - no viscosity) from \cite{EMh}.
         \textbf{Center}: therefore, also Rabi cycle needs hydrodynamical analog, tank of liquid with standing wave seems analogous to laser resonator. Being semi-transparent/semi-open in one direction (also in Fig. \ref{coupling}), it should cause left-right swimming of a probe particle outside, through exchange of momentum acting in both directions, using positive/negative pressure. In CPT perspective it looks the same, also with energy/momentum exchange in both directions. Another analogy is fan/marine propeller exciting resonator (by swirl-like wave carrying energy+momentum+angular momentum like photon), or deexicitng it if reversing rotation direction. 
        \textbf{Bottom}: while the above bidirectional source is resonator extending to external targets, coupling with them through standing wave e.g. for Rabi oscillations, there are also unidirectional sources. In this analogy, synchrotron radiation is kind of water stream allowing only unidirectional momentum transfer. Optical isolator analog could be some unidirectional valve, analog of ring laser: a pump enforcing unidirectional circulation - with reversed circulation in CPT perspective. Radiation pressure is a vector: can be also negative~\cite{neg1,neg2}, suggesting to build more symmetric and powerful two-way microfluidic, photonic computers as in Fig. \ref{2wqc}.}
        \label{rabi}
\end{figure}

\section{CPT symmetry in photonics arguments}
Let us gather here some arguments for the expected $(\leftarrow)$ possibility mainly based on CPT symmetry. However, as in Fig. \ref{test}, it seems to also imply textbook $(\rightarrow)$ case, hence both can be true - hence practically this is a test of existence of $(\leftarrow)$ effect.
\subsection{Bidirectional and forward/backward beams}

\textbf{Stimulated emission-absorption} equations are at heart of lasers, governing deexcitation-excitation e.g. in laser active medium. For $N_1$ atoms in the ground state, $N_2$ excited:
\be\textrm{stimulated emission:}\ \frac{\partial N_2}{\partial t}=-\frac{\partial N_1}{\partial t}=-B_{21}\,\rho(\nu)\, N_2\label{ee}\ee
\be\textrm{and absorption:}\ \frac{\partial N_2}{\partial t}=-\frac{\partial N_1}{\partial t}=B_{12}\,\rho(\nu)\, N_1 \label{ae}\ee 
\noindent where $B_{12}=B_{21}$ are (CPT symmetric) Einstein's coefficients~\cite{Einstein}, $\rho(\nu)$ is radiation density of the incident field at frequency $\nu$, corresponding to transition between the two considered states: $E_2-E_1=h\nu$.

While both are assumed to act on active medium inside laser, naively only absorption equation acts on external target - for example STED laser demonstrates it is not true, acting e.g. by diode laser with stimulated emission equation on external target. However, this equation is negligible for $N_2\approx 0$ - requires initial excitation, while unprepared target is close to the ground state.

As in Fig. \ref{dirs}, the situation changes for example for \textbf{ring laser} with optical isolator - using Faraday effect~\cite{faraday} to allow photon circulation only in one direction. Such photons could leave the laser and be absorbed by a (forward) external target. 

Looking at such unidirectional ring laser from CPT perspective, photon trajectory would be reversed, causing excitation of different (backward) target - increasing $N_2$ toward minus time, hence decreasing $N_2$ in standard time perspective - as in stimulated emission. Therefore, this CPT symmetry switches the two equations (if reversed bias) - in contrast e.g. to diode lasers in STED microscopy, seems only one equation at the time acts on the  forward/backward targets - allowing to split beams into intuitively \textbf{forward/backward beam} acting with \textbf{separated absorption/stimulated emission equation} alone, combined into \textbf{bidirectional beam} in standard e.g. diode laser.

Another example of laser sources switching targets in CPT perspective are based on \textbf{synchrotron radiation} like free electron laser/synchrotron/wiggler/undulator - while there is usually focus only on their forward target, in CPT perspective electron trajectories would be reversed - causing excitation of backward target in the opposite direction, hence should reduce its $N_2$ in standard time perspective, as in stimulated emission.

As noticed in Fig. \ref{test}, forward/backward beams might have completely different \textbf{transparency} as they act only on $N_1$ deexcited / $N_2$ excited atoms. In other words, while textbook view on stimulated emission means amplification of forward beam by going through excited medium, its CPT analog is amplification of backward beam by going through deexcited medium (as in this Figure). Therefore, as everyday matter usually has $N_2/N\approx 0$, backward beam might have much better penetration, allowing for novel applications. However, coupling standing wave might need to reach the target, what could essentially weaken such effect. 

While application of CPT symmetry seems abstract, general relativity in theory allows a more tangible possibility useful for intuitions - in theory allowing Klein-bottle-like nonorientable wormholes~\cite{nonorientable} applying P or T symmetry to e.g. a rocket travelling through which - for T symmetry switching absorption/stimulated emission inside, forward/backward beams.

\begin{figure}[t!]
    \centering
        \includegraphics[width=9cm]{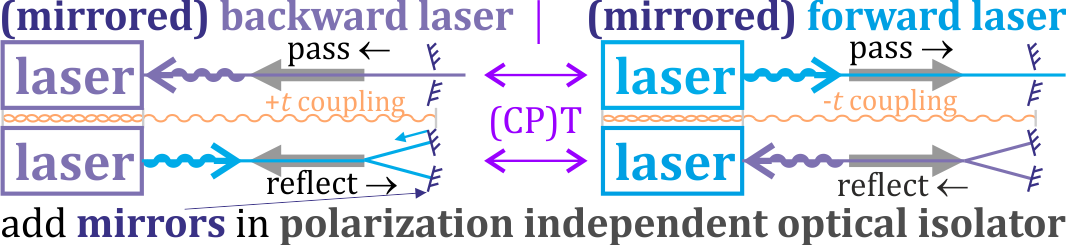}
        \caption{While most laboratory lasers have built-in (forward) optical isolator - can be referred as \textbf{forward laser} (only pushing photons), especially for Rabi cycle as in Fig. \ref{coupling} it is worth to distinguish (\textbf{bidirectional} or) just \textbf{laser} having removed optical isolator. For various mentioned applications it is also worth to consider \textbf{backward laser} as CPT transformed forward laser: with reversed bias and optical isolator, should only pull photons with negative radiation pressure. Polarization independent optical isolators usually change angle for the opposite photon direction, allowing to add mirrors inside to reflect back such photons (e.g. extending laser resonator toward one time direction) - \textbf{mirrored backward laser} with isolator containing such mirrors should be much more energy efficient, might also improve forward laser as having lower energy loss toward $-t$. Angle, polarization, path for to reflect such reverse photons would need optimization, similar effect could be realized with circulator, ideally would be getting such effect with metamaterials.   }
        \label{bl}
\end{figure}

Spontaneous emission is very different and seems asymmetric in time, but as for entropy asymmetry, it should be a consequence of concrete solution we live in. From Feynman diagram perspective, such exchanged photon still couples between e.g. two electrons exchanging photon - with practically random target for spontaneous emission. As in further Fig. \ref{asym}, this asymmetry, also of circulating electrons losing energy, seems to come from now having more absorbers in the futures, than emitters in the past - what might weaken in the future. This asymmetry seems also crucial for two scenarios at the bottom of Fig. \ref{test} - forward beam is likely to cause deexcitation as there are many absorbing targets in the future, what is not currently true for backward beam: should nearly only pull photons from excited target.

\subsection{Optical cavity and delocalized standing EM wave coupling}
In vacuum Rabi cycle~\cite{cavity} e.g. a single atom periodically exchanges energy with optical cavity - cycling energy between atomic orbital, and standing EM wave of cavity, with energy density more localized in the former case. This difference is similar to between laser and radio communication - the former uses localized optical photons travelling in one direction, the latter delocalized e.g. spherical EM waves travelling in all directions.

While both are usually described with Fock space, this is abstract effective description of pertubative approximation. Like "apple + apple = 2 apples" correct algebra, for deeper dependencies requiring to ask about structure, field configuration in non-perturbative picture, here mainly of electromagnetic fields - again with higher energy density when localized in atom e.g. during vacuum Rabi oscillations.

Being different types of EM field configurations, their interactions might also differ, for example with sophisticated devices like optical isolator - specialized to block localized optical photons travelling in one direction, but not necessarily delocalized standing EM waves for coupling - what should be tested, e.g. with setting as in Fig. \ref{coupling}. As in Fig. \ref{bl}, some optical isolators could be enhanced with mirrors to be reflective instead of blocking in the opposite direction, useful e.g. to enclose resonator of potential backward lasers: only amplifying target deexcitation, e.g. for radiotherapy as in Fig. \ref{radiotherapy}.

\subsection{Rabi cycle as coupling, its asymmetrization}
Generally, in Rabi cycle as in Fig. \ref{coupling}, analogously to coupled pendula (e.g. \cite{crabi1,crabi2,crabi3,crabi4}), there is observed coupling of two resonators periodically exchanging energy and momentum in both directions - in practice using mainly lasers as pumped directional resonators, atoms as tiny isotropic resonators, or cavities leading to standing waves/eigenmodes. 

As electromagnetism and hydrodynamics are governed by nearly the same equations, it is also valuable to think about Rabi cycle from hydrodynamics perspective as in Fig. \ref{rabi}: laser as standing wave extending outside resonator due to semi-transparent wall, exchanging energy and momentum with external objects in both directions - with positive and negative pressure, both exist also for radiation~\cite{neg1}.

\begin{figure}[t!]
    \centering
        \includegraphics[width=9cm]{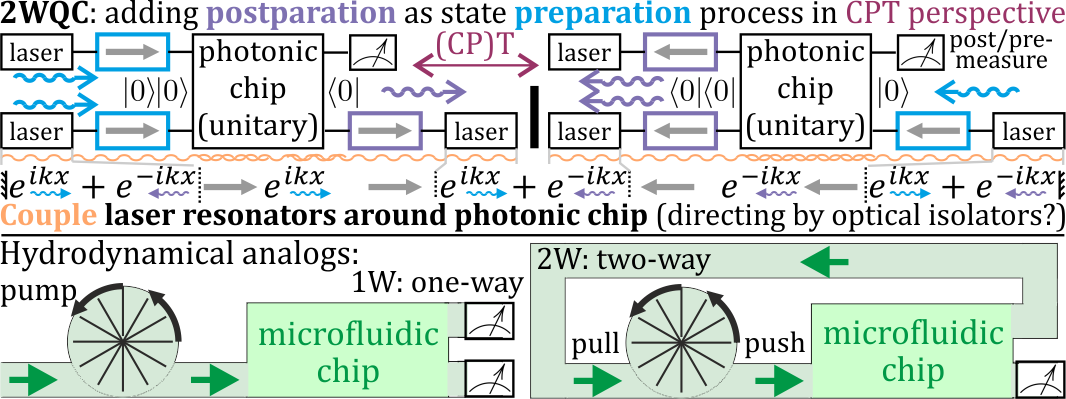}
        \caption{Original motivation: (example of) two-way quantum computer (2WQC)~\cite{2WQC} (and its classical hydrodynamical analog) as adding a process which in CPT perspective becomes the original state preparation process. For photonic QC state preparation is usually performed by laser impulse, suggesting to just add 'reversed' impulse: delayed by propagation time, using reversed bias, and behind backward placed optical isolator - this way from CPT perspective photon trajectories should be reversed, like improving two-way flow control in hydrodynamical analog. It can be also viewed as coupling of two resonators, with photonic chip on the way. In comparison to standard 1WQC, in theory 2WQC allows for up to exponential speedup e.g. to solve postBQP~\cite{postBQP}, NP problems, also better stability and error correction~\cite{Grover,3sat2wqc}. Such classical 2WCC photonic computer on continuous beam or impulses seems also worth considering. }
        \label{2wqc}
\end{figure}

Such bidirectional energy exchange for e.g. laser-atoms coupling is mainly realized by optical photons, hence we can view standard e.g. diode laser as bidirectional beam - causing both excitation of target, but also deexcitation - e.g. in Rabi cycle, STED microscopy, or backward ASE of EDFA.s

Being able to introduce asymmetry in this coupling could lead to new applications. For example free electron laser (FEL) does not have such resonator coupling, 
CPT symmetry suggests that looking at such synchrotron radiation beam from CPT perspective, photon beam created by reversed electron trajectory should act as backward beam - causing only target deexctiation, acting only with stimulated emission equation.

\begin{figure}[t!]
    \centering
        \includegraphics[width=9cm]{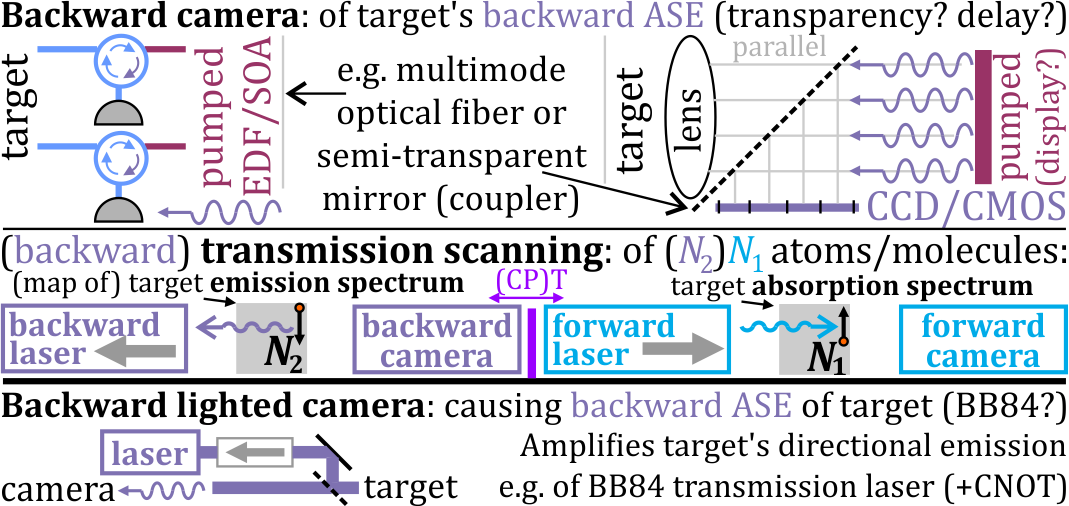}
        \caption{\textbf{Top}: Backward camera/detector observing negative radiation pressure/stimulated emission from the target (or added backward laser), crucial for some further discussed applications. Detector can be e.g. based on (multi-mode) fiber with couplers/circulators as in \cite{bASE} and Fig. \ref{erb}. 
        For camera there can be used semitransparent mirror and CCD/CMOS - both need some continuously pumped e.g. 3-level medium (maybe from some display technology). This approach might be also possible for different EM frequencies, using e.g. antenna with excited LC resonator, or maybe stimulating emission from accelerating electrons e.g. of betatron, synchrotron, vacuum tube lamp in magnetic field. Also light diode in forward bias could be used - monitoring increased current if negative radiation pressure acts on it.
        Alternatively one could monitor spontaneous emission of the pumped medium, e.g. with camera behind. There could be added optical isolator to remove incoming photons. As in Fig. \ref{CAT}, pairing backward detector/camera with pointing backward laser could allow for transmission scanning for up to 3D maps of excited molecules between (e.g. medical, material science, geological scanning) - like in standard transmission scanning, but of emission spectrum instead of absorption.
        \textbf{Bottom}: Backward lighted camera: applying backward ASE effect on the target to amplify its emission in chosen spectrum toward camera (e.g. thermal also cooling), or on laser medium used to send quantum cryptography protocols like BB84, to increase its number of produced photons and steal/read some - e.g. going through control of CNOT gate to work on single photons.  }
        \label{camera}
\end{figure}

\section{Some potential new applications}
The above considerations could lead to many new applications, especially if generating backward beam of negative radiation pressure acting on target only with stimulated emission equation. Optimistically it can be obtained by reverse biased diode/laser (without or with reversed optical isolator). Otherwise, as in Fig. \ref{synch}, \ref{dirs}, should be possible with antennas and synchrotron radiation, e.g. betatron, synchrotron, free electron laser, or bremsstrahlung e.g. in X-ray lamp in strong magnetic field. For example:

\begin{itemize}
 \item Backward beam, e.g. from synchrotron radiation, could allow for CPT analog of state preparation of quantum computers 
     for 2WQC~\cite{2WQC} as in Fig. \ref{2wqc}, in theory allowing for up to exponential speedup (solve postBQP~\cite{postBQP}) and better error correction~(\cite{Grover, 3sat2wqc}). 
 \item If we could send information this way, there could be e.g. built time-loop computers~(Section V of \cite{my}) for example using backward beam to send output of chip back to its original input with negative time delay, also in theory allowing to solve NP problems. Connecting such multiple negative delay transmitters e.g. into a cycle, could allow to also reach macroscopic negative time differences. 
 \item Improvements of current technologies, like STED microscope not photobleaching the sample e.g. by reverse biased laser  backward beam causing stimulated emission alone.  
  \item Backward beam in theory should be transparent for $N_2=0$ medium, what could allow for much better penetration than forward (transparent for $N_1=0$) - e.g. for radiotherapy as in Fig. \ref{radiotherapy}, backward camera as in Fig. \ref{camera} and various scanners e.g. emission CT for 3D map of $N_2$ atoms/molecules as in Fig. \ref{CAT}, communication through obstacles, etc.
 \item Antenna backward beam could direct energy of transmitter toward user, e.g. to amplify, eavesdrop or for jamming.
 \item Backward beam to stimulate specific atomic, chemical or nuclear processes: releasing emission spectrum photons of chosen energy (and polarization), e.g. for chemical/technological processes,  energy production, or medical therapies as in Fig. \ref{radiotherapy} - e.g. locally reduce lifetime of crucial for metabolism molecule like NADH or FADH to starve cancer tissue and release this energy as thermal, or inhibit biochemical processes by amplifying deexcitation of crucial molecules, or injected contrast medium, for example with multiple rotating beams intersecting in cancer tissue, especially if better transparency is confirmed.  
     \begin{figure}[t!]
    \centering
        \includegraphics[width=9cm]{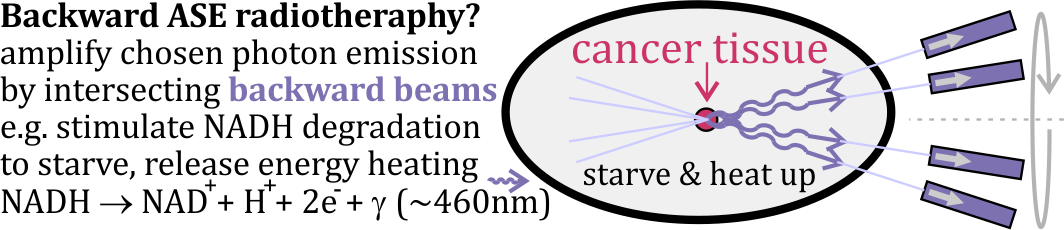}
        \caption{If being able to separate absorption and stimulated emission equations, for example with CPT analog of synchrotron radiation: beam resulting from reversed electron trajectories as in Fig. \ref{dirs}, or maybe reverse biased laser with backward placed optical isolator as in Fig. \ref{bl}, one of many new applications could be radiotherapy trying to locally stimulate emission of photons of chosen energy (and polarization). For example to try to locally speedup degradation of energy carrying molecules like NADH (nicotinamide adenine dinucleotide) or FADH crucial for cell energy metabolism - to starve cancer tissue, and heat up by releasing this energy. Or other autofluorescent molecules with emission spectrum (lists e.g. in \cite{lum1,lum2,lum3}) to locally influence metabolism. Intersecting multiple such beams, or e.g. one beam spread in 1D by two cylindrical lenses, additionally rotating them for fixed focus intersection point, we could make the (usually nonlinear) local effect orders of magnitude stronger then for intermediate tissues, to make it negligible outside of targeted tissue. Other issues could be also approached this way, e.g. applying contrast agent as in photodynamic therapy~\cite{photodynamic}, or trying to disable toxic molecules (like $\sim$1280nm for singlet oxygen~\cite{oxy}), influence local neural activity, inhibit chosen biochemical pathways by stimulating deexctiation of crucial molecules e.g. of pathogens, etc.  }
        \label{radiotherapy}
\end{figure}

  \begin{figure}[t!]
    \centering
        \includegraphics[width=9cm]{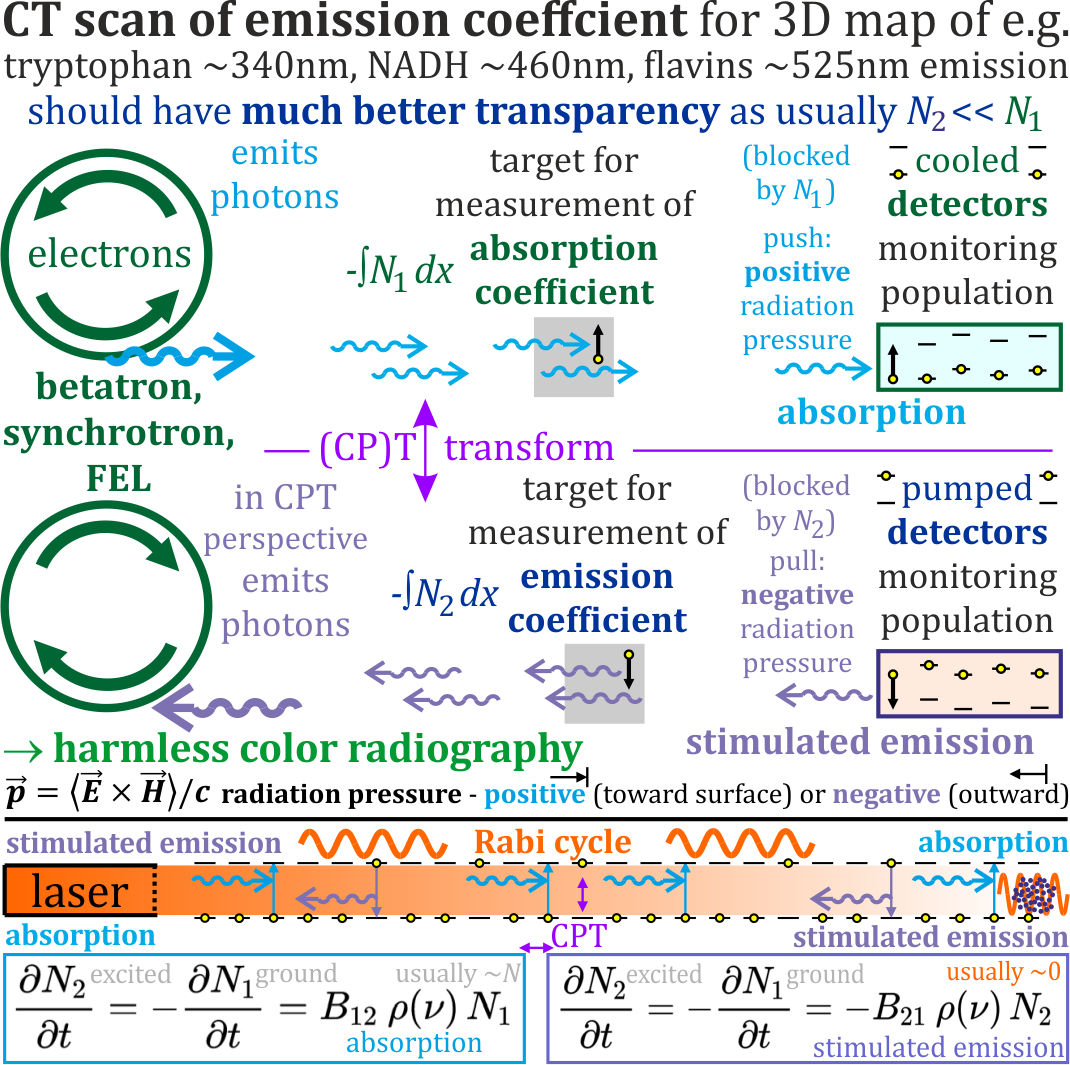}
        \caption{Simplified diagram for measurement of (integrated over light path) absorption coefficient, which can be extended to 3D mapping in \href{https://en.wikipedia.org/wiki/CT\_scan}{CT scanner} (computer tomography) from 2D projections in multiple angles, reconstructed e.g. by Radon transform~\cite{radon}. Above CPT symmetric setting suggests how to analogously measure emission coefficients, what could be extended to its 2D projections, videos (e.g. during surgery), and finally CT scanning for 3D maps. For medical applications e.g. of autofluorescent molecules (gaining energy from cell biochemistry) to make 3D maps of e.g. tryptophan ($\sim$340nm), collagen ($\sim$360nm), elastin ($\sim$410nm),  NADH ($\sim$460nm), flavins ($\sim$525nm)~\cite{lum1}, or some contrast agent - especially if possible to separate stimulated emission, and indeed being blocked only by molecules emitting in the used frequency (better transparency). Using optimized multiple frequencies, could allow for noninvasive detailed (high spatial and temporal resolution) 2D color photos/videos, or 3D maps of tissues, their metabolism with distinguished pathologies like cancer, maybe of local neural activity - multispectral/color, harmless (without ionizing radiation). The source could be synchrotron radiation, maybe reverse biased diode/laser. The detector could be backward camera as in Fig. \ref{camera}: pumped surface (e.g. screen) with monitored population, or adding e.g. half-silvered mirror and standard detector, or maybe forward biased diode with monitored current. }
        \label{CAT}
\end{figure}

  \begin{figure}[t!]
    \centering
        \includegraphics[width=9cm]{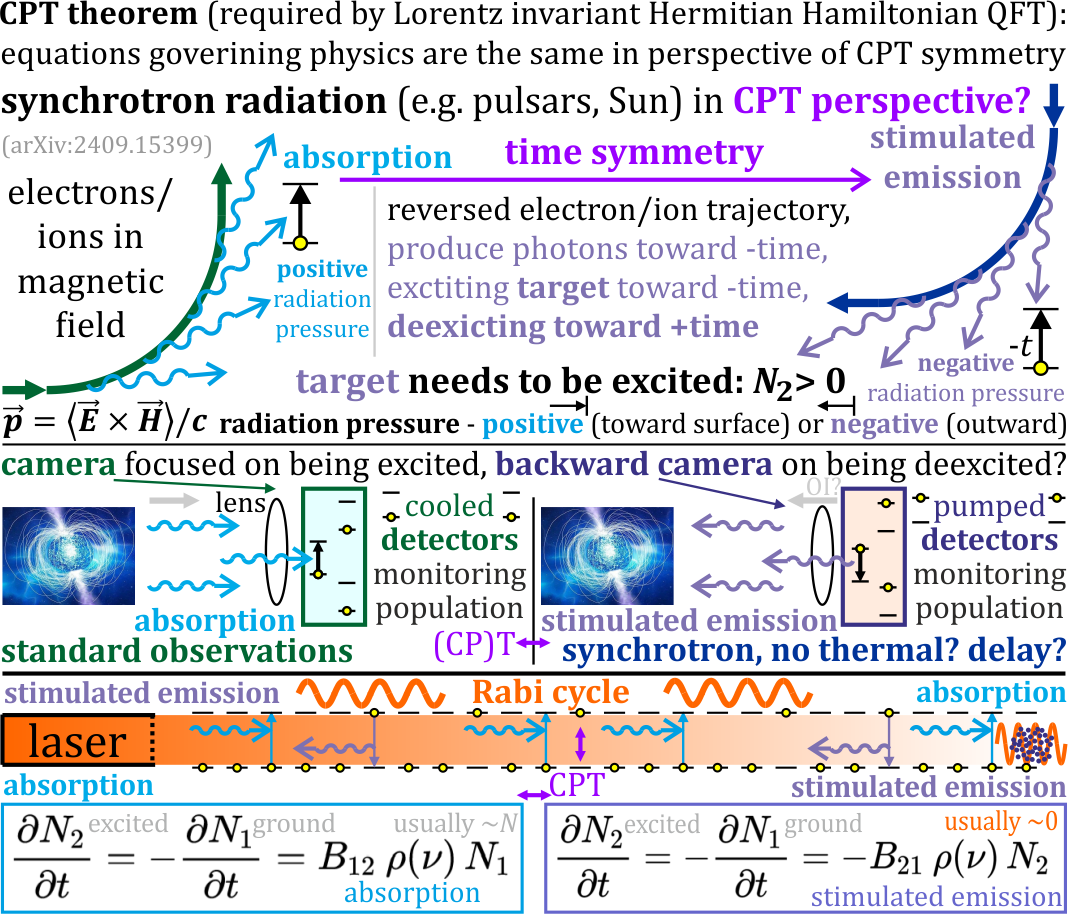}
        \caption{Potential applications of proposed backward camera to obtain complementary information from astronomical objects, e.g. might allow detection/observation of black holes as in Fig. \ref{BH}, \ref{BHdet}, \ref{chirp}, maybe also finding some novel currently unexpected objects, and mapping emission asymmetry for cosmological studies. Many of such objects e.g. \href{https://en.wikipedia.org/wiki/Pulsar}{pulsars} (image source) have accelerating charged particles, what leads to synchrotron radiation - e.g. causing excitation in cooled detector of standard telescopes. In CPT perspective of this scenario, there is still accelerating charge (in opposite direction) - should emit photons, causing excitation now toward our negative time, which is deexcitation toward positive time. Therefore, such process required by CPT symmetry would need excited telescope sensor - usually they are cooled instead, what might be the reason for not observing such effect, could lead to worth searching consequences like cooled halos around such pulsars. To have a chance to observe such effect, we would need to pump the sensor/detector (instead of standard cooling), again monitoring its population, or maybe using semi-transparent mirror as in Fig. \ref{camera} with various photon sources (e.g. antenna, synchrotron, X-ray lamp, forward bias diode with monitored current), maybe also include some optical isolator to allow only outgoing photons. If successful, such telescope should see (reversed) synchrotron, but not thermal radiation - e.g. allowing to separate them when combining with standard picture, also for observations/diagnostics/early warnings for our Sun~\cite{sun}. }
        \label{astro}
\end{figure}

  \begin{figure}[t!]
    \centering
        \includegraphics[width=9cm]{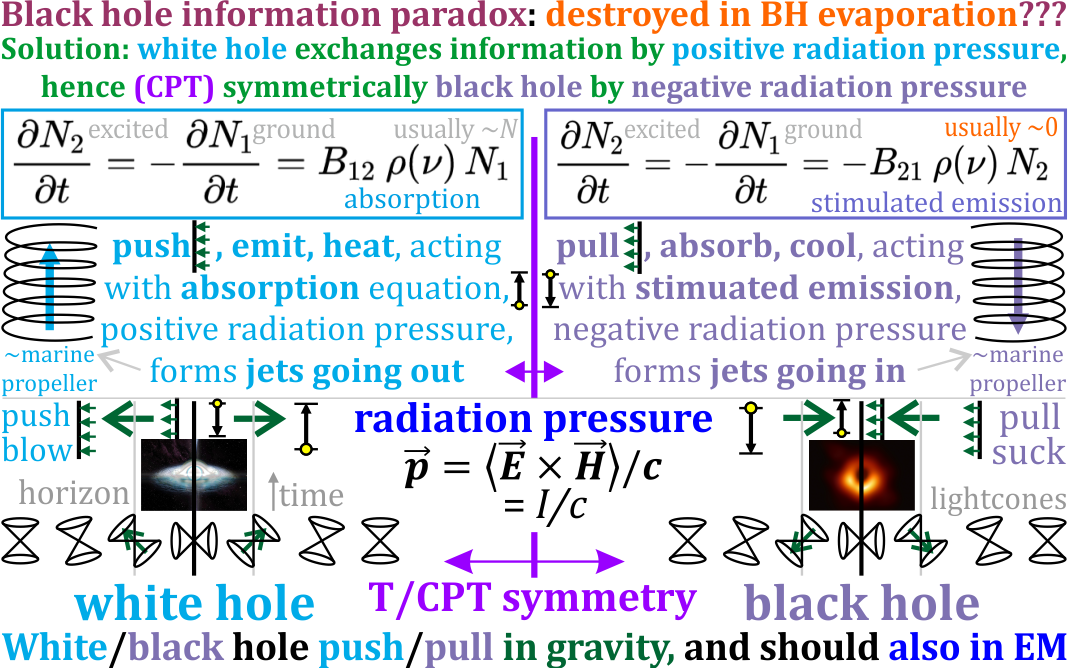}
        \caption{Applying T/CPT symmetry to our \href{https://en.wikipedia.org/wiki/Black_hole}{black hole} (running movie backward), it is a  \href{https://en.wikipedia.org/wiki/White_hole}{white hole}, and they actively emit - acting with positive radiation pressure $\vec{p}=\langle \vec{E}\times \vec{H}\rangle/c$, heating matter around, what returning to our time perspective means that black hole should also actively cool matter around with negative radiation pressure. Hence white/black holes should repel/attract not only gravitationally, but also EM by radiation pressure - continuously interacting/exchanging information with the surrounding in symmetric way, this way allowing unitary evolution, avoiding information paradox~\cite{BHIP} symmetrically for both white and black hole. It also would help forming dynamical outward/inward jets for both, and indeed they turn out very dynamical~\cite{BHvar}. As white holes would heat matter around, black hole should symmetrically cool it - with the same spectrum, but replacing absorption with stimulated emission equation, and indeed matter~\cite{jets}, especially electrons~\cite{twotemp} around turn out colder than expected, it also might prevent observation of IMBH (intermediate mass black holes). Such pull not only in gravity, but also in EM, would additionally increase growth of black hole, which indeed turns our faster than expected~\cite{growth}, generally might be faster than Hawking radiation - suggesting to reconsider their future evolution. While we might already observe indirect consequences of such stimulated emission caused by black hole, backward telescope like in Fig. \ref{camera}, \ref{astro}, \ref{BHdet} might also observe it directly - also below horizon, symmetrically to white hole observation. Black holes are hypothesized to have negative pressure~\cite{nBH}, negative temperature pressure ~\cite{npBH}, which should also lead to radiation pressure through reversed emission/absorbtion tendency as in Fig. \ref{negtemp}. }
        \label{BH}
\end{figure}  

 \begin{figure}[t!]
    \centering
        \includegraphics[width=9cm]{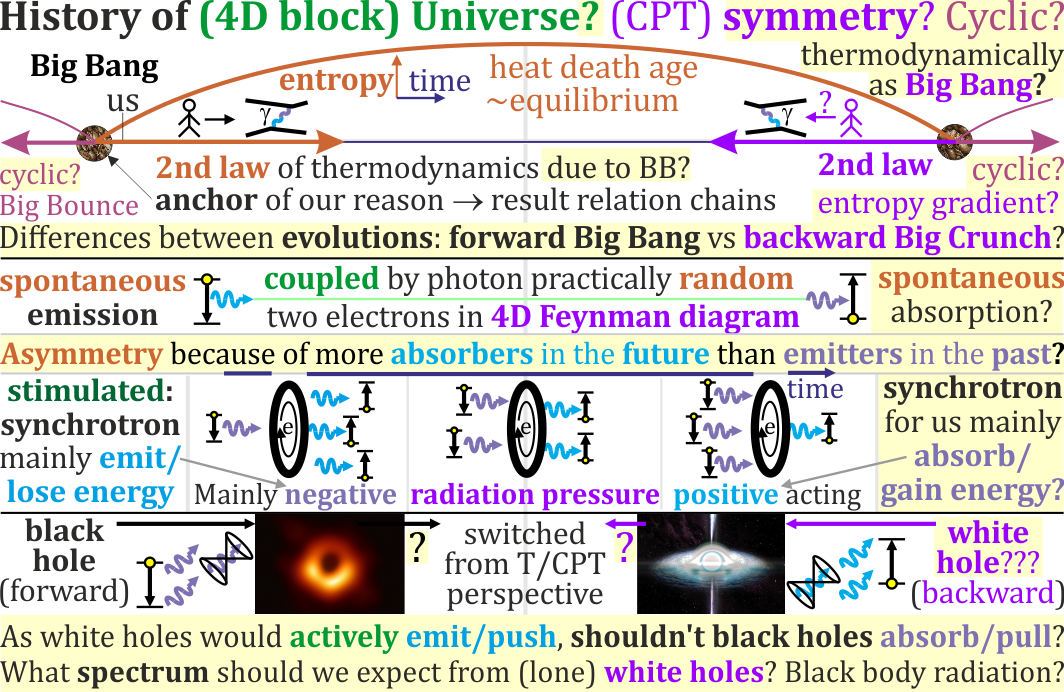}
        \caption{Many related questions (yellow) regarding T/CPT symmetry and its violation in solution like \textbf{entropy growth}, assuming Big Crunch scenario, which seems more likely after recent DESI results of weakening dark energy~(\cite{DESI,lifespan}) and revisited supernova data~\cite{supernova}. Thermodynamical conditions of Big Crunch seem very similar to of Big Bang, hence should be also entropy - requiring to finally reverse entropy gradient in such 4D Einstein's block Universe~\cite{block} solution we are travelling through, making Big Crunch similar to time reversed Big Bang. There is also \textbf{emission asymmetry}, the simplest for circulating electron - now it loses energy, while from CPT perspective it gains energy. While temperature should be symmetric, a natural explanation seems because there are now more absorbers in our future than emitters in our past - which can be switched e.g. by CPT symmetry, maybe near Big Crunch (switching times), and practically is achieved in tabletop particle accelerators~\cite{tabletop} adding emitters in laser. Backward camera could map anisotropy of this emission asymmetry to bring valuable hints for cosmology, analogously to COBE/WMAP satellites. Finally, there is \textbf{tendency for black holes formation} after Big Bang, suggesting symmetric tendency for white holes before Big Crunch - if the former could survive till the end, maybe the latter could also till our times - suggesting to consider possibility that some lone stars could turn out white holes, requiring to understand what spectrum should we expect, e.g. by negating $\rho(\nu)$ density measured for black holes, maybe also possible to observe by gravitational waves as in Fig. \ref{chirp}.       
         }
        \label{asym}
\end{figure}  

 \begin{figure}[t!]
    \centering
        \includegraphics[width=9cm]{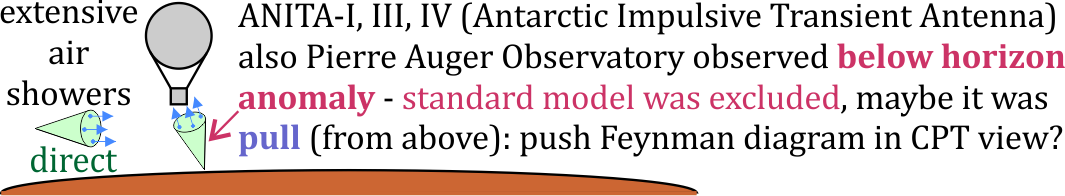}
        \caption{ANITA experiment~\cite{anita} uses array of radio antennas suspended from a helium balloon. It observes e.g. extensive air showers caused e.g. by ultra-high energy neutrinos ($>10^{18}$ eV). However, some of the showers seem to come from below Earth - such events are referred as ANITA below horizon anomaly, observed by ANITA-I, III, IV missions, also recently by Pierre Auger Observatory~\cite{auger}. Standard model candidates crossing Earth are claimed to be excluded~\cite{SManita}. Therefore, it might be worth to consider \textbf{pull hypothesis} - that they were caused by an astronomical scenario which in CPT perspective corresponds to Feynman diagram for symmetric push scenario.       
        }
        \label{anita}
\end{figure}  

 \begin{figure}[t!]
    \centering
        \includegraphics[width=9cm]{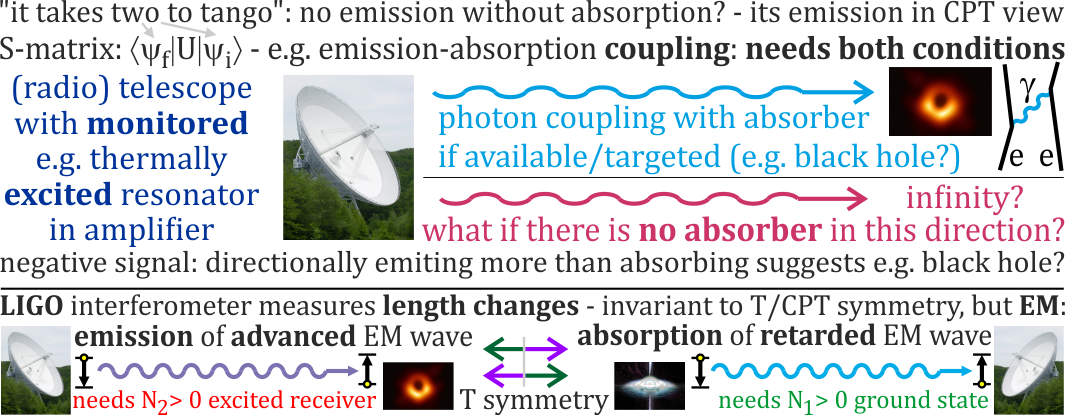}
        \caption{Potential observation of black holes with (radio)telescope. From CPT perspective: running movie backward, this would be white hole emitting photons (from below horizon) which could be absorbed by telescope - what in our time perspective means emission by telescope stimulated by black hole, however, it requires earlier excitation of telescope sensor. While usually radio telescope amplifiers are cooled, avoiding it or actively exciting resonator inside, in theory this antenna could also emit directionally. However, used frequencies are often difficult to absorb in cosmos, without specific absorber might go to infinity, what in CPT perspective would mean such EM wave was never emitted, hence emission should be forbidden if not pointing absorber, like black hole - suggesting novel way to detect them, and observe also below horizon, symmetrically to white hole emitting from below horizon. Another view is by $\langle \psi_f |U|\psi_i \rangle$ \href{https://en.wikipedia.org/wiki/S-matrix\#Interaction_picture}{S-matrix}: with amplitude/probability of photon depending both on emitter in $\psi_i$, but also absorber in $\psi_f$, they are switched in CPT view - without absorber, in CPT view there is no emitter. In contrast, LIGO measures lengths, which are T/CPT symmetric: should be able to see both retarded and advanced as in Fig. \ref{chirp}.      }
        \label{BHdet}
\end{figure}  

 \begin{figure}[t!]
    \centering
        \includegraphics[width=9cm]{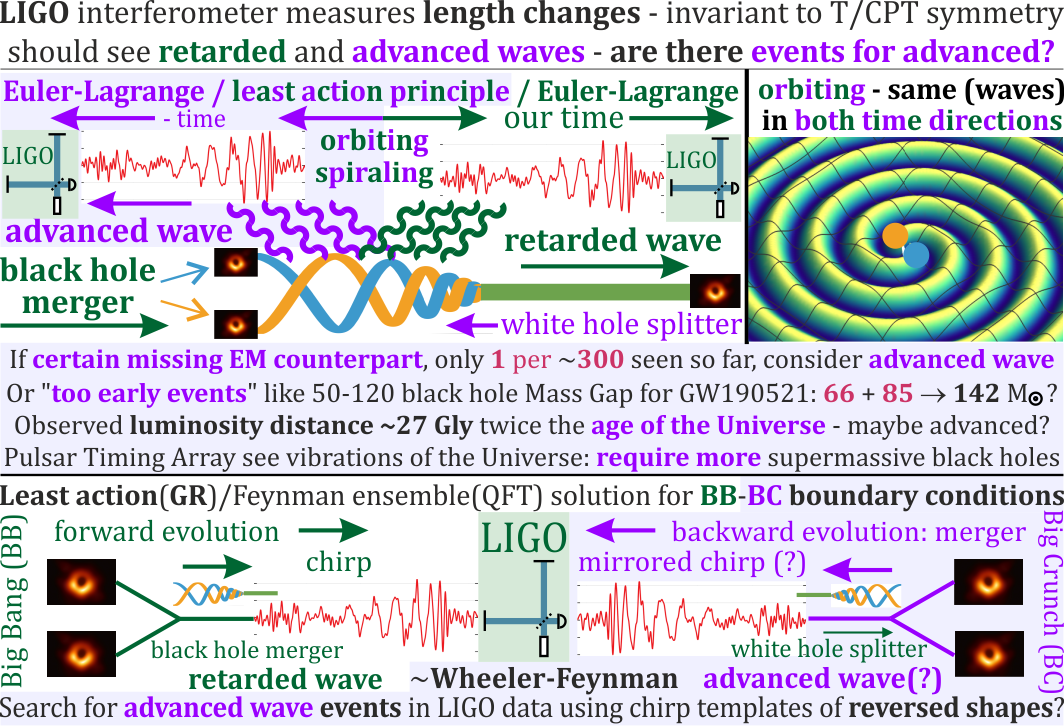}
        \caption{\href{https://en.wikipedia.org/wiki/Wheeler\%E2\%80\%93Feynman_absorber_theory}{Wheeler-Feynman absorber theory} requires both standard retarded EM waves, but also symmetrically advanced, and they are switched in T/CPT view. As general relativity is rather solved by the least action principle, it should be also true for gravitational waves. \href{https://en.wikipedia.org/wiki/Gravitational_wave\#LIGO_and_Virgo_observations}{LIGO} measures lengths, which are T/CPT invariant, so should see both retarded and advanced waves - the big question is if there exist events for the latter.
        \textbf{Top}: observed waves from e.g. black hole mergers come mainly from orbiting of two masses, what is practically the same from time symmetry view - so should emit gravitational waves in both perspectives. Using forward Euler-Lagrange evolution would give retarded waves, but we could also use it backward, or should rather solve with the least action principle - both rather requiring also symmetric advanced wave, of similar chirp shapes at least in orbiting regime. As in the \href{https://en.wikipedia.org/wiki/List_of_gravitational_wave_observations}
        {list of gravitational wave observations} there was still observed only single EM counterpart from around 300 observed events, $\approx 10$ involving neutron stars, some of them could turn out advanced waves, especially that there should be more opportunities in our future than our past. Some suggestions for advanced waves could be events claimed to happen earlier than possible like GW190521~\cite{tooearly}, or extremely large luminosity distances - now observed up to twice the age of the Universe. Additionally, MeerKAT pulsar timing array observe vibrations of the Universe~\cite{vibration} which seem to require more orbiting supermassive black holes than expected - including advanced could help.
        \textbf{Bottom}: another hypothetical scenario, as in Fig. \ref{asym} assuming Big Crunch - from which backward evolution seems similar to forward from Big Bang, hence both should have tendency to form black holes, which mergers should have reversed shape chirps - suggesting to search LIGO historical data for dataset of chirp templates with applied time symmetry. Another candidate for advanced wave might core-collapse events, forming pulling wave.
        }
        \label{chirp}
\end{figure}  

 \item Backward camera as in Fig. \ref{camera} maybe exploiting transparency and/or opposite delay e.g. for astronomical observations (of e.g. synchrotron radiation of pulsars,  our Sun~\cite{sun}, maybe black holes as in Fig. \ref{astro}, \ref{BH})- focused on target's negative radiation pressure (or transmission of added backward beam), while standard (forward) camera is focused on positive. It would replace CMOS/CCD matrix behind the lens, with matrix/surface of continuously excited e.g. approximately 3-level medium like dye or erbium, plus some monitoring of its population map - e.g. through spontaneous emission, or EIT (electromagnetically induced transparency),  half-silvered mirror and CMOS/CCD, or maybe forward biased diode with monitored current. Pairing with backward beam could allow for various scanning (e.g. medical, material science, geological) allowing (emission) CT scanning for 3D maps of $N_2$ excited atoms/molecules between them as in Fig. \ref{CAT}. To observe negative radiation pressure for non-visible spectrum, there could be used antennas with excited resonator, or maybe X-ray lamp, or electrons travelling e.g. in synchrotron/vaccum tube lamp as having potential to emit such photons.   
 \item Backward beam for attacks on quantum cryptographic protocols like BB84~\cite{bb84} - use external laser to additionally stimulate emission inside communication laser (as backward ASE) to steal/intercept such additional photons, which could be read e.g. by passing through control qubit of CNOT gate - affecting the second qubit to be measured.
 \item Backward-lighted camera as in Fig. \ref{camera}: with added backward beam for observations actively increasing deexcitation rate of excited objects, e.g. for thermal imaging, of contrast agent, autofluorescent biomolecules, etc.
 \item Backward beam for 3D printing/reversed photolithography: excite material and precisely stimulate deexcitation (can be in 3D as for radiotherapy) - controlling position, time, frequency, also polarization to choose one of multiple possible transitions e.g. of high energy for local cooling.
 \item Backward beam for optical cooling and pulling, directly acting with negative radiation pressure.
 \item Synchrotron sources usually have much higher photon energy, its backward beam could e.g. increase probability of selected nuclear decays/transitions: producing photons of chosen energy. Maybe also increasing probability of other particle events/Feynman diagrams: which CPT transform requires such photons incoming from e.g. synchrotron.
 \item Other potential astronomical applications are e.g. COBE/WMAP-like mapping of cosmic anisotropy for emission asymmetry (of absorbers, maybe Big Crunch) by backward camera as in Fig. \ref{asym}. Or as in Fig. \ref{anita}, for ANITA anomaly  it might be worth to consider that it was some pull instead of pushing, by Feynman diagram of process which in CPT view is pushing. Also for gravitational waves there should be similar effects, e.g. LIGO just measuring lengths should see both retarded and advanced waves as in Fig. \ref{chirp}, which seems worth being included in considerations, maybe allowing to find some new events e.g. by searching for chirps of time-reversed shape.
\end{itemize}

\section{Conclusions and further work}
There were proposed various ways to search for negative radiation pressure, CPT analog of emission, the safest with synchrotron radiation, which is planned to be tested in some future. The further work depends on the result:
\begin{itemize}
\item obtaining it as required by CPT symmetry should lead to many new applications to explore, like the ones listed above,
\item its lack would indicate macroscopic violation of CPT symmetry, requiring further investigations of its details.
\end{itemize}

\bibliographystyle{IEEEtran}
\bibliography{cites}
\end{document}